\documentclass[%
 aip,
 amsmath,amssymb,
 reprint,%
]{revtex4-2}

\usepackage{graphicx}
\usepackage{dcolumn}
\usepackage{bm}

\usepackage[percent]{overpic}

\usepackage[utf8]{inputenc}
\usepackage[T1]{fontenc}
\usepackage{mathptmx}
\usepackage{color}

\newcommand\R{\mathbb{R}}
\newcommand\x{\mathbf{x}}
\newcommand\f{\mathbf{f}}
\newcommand\ddt{\frac{\mathrm{d}}{\mathrm{d}t}}

\begin{document}

\preprint{AIP/123-QED}

\title[A study of the double pendulum using polynomial optimization]{A study of the double pendulum using polynomial optimization}

\author{J. P. Parker}
\email{jeremy.parker@epfl.ch}
\affiliation{
Emergent Complexity in Physical Systems Laboratory (ECPS), 
École Polytechnique Fédérale de Lausanne, 1015 Lausanne, Switzerland
}
\author{D. Goluskin}
\affiliation{
Department of Mathematics and Statistics,
University of Victoria, Victoria, BC V8P 5C2, Canada
}
\author{G. M. Vasil}
\affiliation{
School of Mathematics and Statistics,
University of Sydney, Sydney, NSW 2006, Australia}

\date{\today}

\begin{abstract}
In dynamical systems governed by differential equations, a guarantee that trajectories emanating from a given set of initial conditions do not enter another given set can be obtained by constructing a barrier function that satisfies certain inequalities on phase space. Often these inequalities amount to nonnegativity of polynomials and can be enforced using sum-of-squares conditions, in which case barrier functions can be constructed computationally using convex optimization over polynomials. To study how well such computations can characterize sets of initial conditions in a chaotic system, we use the undamped double pendulum as an example and ask which stationary initial positions do not lead to flipping of the pendulum within a chosen time window. Computations give semialgebraic sets that are close inner approximations to the fractal set of all such initial positions.
\end{abstract}

\maketitle

\begin{quotation}
The most common computational approach to studying chaotic trajectories of differential equations is to numerically integrate in time from single initial conditions. This paper illustrates a complementary approach where time evolution of whole sets of initial conditions is studied by solving polynomial optimization problems subject to sum-of-squares constraints.
\end{quotation}

\section{\label{sec:introduction}Introduction}

Although dynamics of chaotic ordinary differential equations (ODEs) cannot be characterized exactly, numerical integration from a single initial condition may approximate various properties of the resulting trajectory. It can be much more difficult, however, to determine properties of whole sets of trajectories emanating from given sets of initial conditions. One such challenge, which we study here, is to verify that no trajectories from a specified set of initial conditions enter another specified part of phase space within a chosen time window. This type of statement cannot be verified solely by computing a finite number of trajectories. Instead, the forward-time evolution of the initial set must be found exactly or, more likely, approximated by larger containing sets.

The approach we use to show that trajectories from an \emph{initial set} do not later enter an \emph{avoidance set} is to construct a type of scalar-valued function of the phase space variable that is called a \emph{barrier function}. When trajectories are governed by ODEs, the time derivative of such a function along trajectories follows from the governing equations without the trajectories themselves being known. If one is able to construct a barrier function such that it and its time derivative satisfy suitable pointwise constraints, this verifies that the avoidance set is indeed avoided. The constraints placed on barrier functions vary but all work in essentially the same way by implying that no trajectories from the initial set later enter the avoidance set.\cite{Prajna2007, Ames2017, Ahmadi2017, Ames2019} One such choice of constraints is for the barrier function to be positive on the initial set, negative on the avoidance set, and nondecreasing along all trajectories. Barrier functions belong to a family of dynamical systems methods that all rely on constructing auxiliary functions---that is, functions defined on phase space that satisfy various pointwise inequalities, which in turn imply various statements about the dynamics. The best-known type of auxiliary function is the Lyapunov function, which implies stability of an equilibrium point by being positive and decreasing along trajectories everywhere else in phase space. As with Lyapunov functions, it can be very challenging to explicitly construct barrier functions that satisfy the desired inequalities in applications to particular ODEs. However, in the important case where the ODE right-hand side has polynomial dependence on the phase space coordinates, barrier functions can be constructed computationally.

Provided that the governing ODE is polynomial and the barrier function is sought from a finite-dimensional space of polynomials, the inequality constraints that the barrier function must satisfy amount to certain polynomial expressions being pointwise nonnegative on phase space or on subsets thereof. Although deciding pointwise nonnegativity of a polynomial is computationally intractable {unless the degree or number of variables is small,\cite{Parrilo2013a}} nonnegativity can be enforced by stronger but more tractable conditions in which polynomials are required to admit sum-of-squares (SOS) representations. In this way, a polynomial barrier function can be sought computationally by solving a type of polynomial optimization problem called an SOS program. These convex optimization problems have found numerous applications since it was realized two decades ago that they can be reformulated as semidefinite programs and solved using existing algorithms.\cite{Nesterov2000, Parrilo2000, Lasserre2001} One of the areas in which applications of SOS programming are proliferating is the computational construction of various types of auxiliary functions for dynamical systems. In addition to barrier functions \cite{Prajna2007, Ahmadi2017} and Lyapunov functions\cite{Papachristodoulou2002, Anderson2015a, Fuentes2019}, other types of auxiliary functions constructed by SOS computations have provided results that include approximations of basins of attraction{\cite{Henrion2014, Korda2013}}, bounds on infinite-time averages or stochastic expectations \cite{Chernyshenko2014, Fantuzzi2016, Kuntz2016, Goluskin2018, Goluskin2019a, Olson2020}, enclosures of global attractors \cite{Schlosser2020} or bounds thereover\cite{Goluskin2020a}, bounds on extreme events \cite{Fantuzzi2020, Miller2021}, and more. The constraints on auxiliary functions used to bound extreme events in Refs.\ \citenum{Fantuzzi2020} and \citenum{Miller2021} are especially similar to the constraints on barrier functions and may yield similar conclusions in some cases.

Previous examples in which barrier functions have been constructed using SOS computations have had relatively simple dynamics, partly because a common use of barrier functions is to verify avoidance of a set of undesirable states in the presence of a control law,\cite{Ames2019} and controls produce simple dynamics by design. Here, however, we apply the barrier function approach to a chaotic system. As a simple set of initial conditions evolves forward in time in a chaotic system, the set's shape becomes increasingly complicated due to the separation of nearby trajectories. This makes it inherently difficult to construct outer approximations of the time-evolved sets that are not overly conservative{, although advances have been made recently using spatiotemporal decomposition.\citep{Cibulka2022} There exist other methods for constructing outer approximations} by covering the initial set in simple shapes and time integrating using interval arithmetic, \cite{Miyaji2016} but this is an active area of research with its own difficulties and is not our current focus. Barrier functions, on the other hand, do not need to approximate time-evolved sets uniformly well---only well enough to determine that they never intersect the avoidance set. Whether this fact allows barrier functions to give useful statements about chaotic dynamics is what the present work aims to determine.

The example we study here is the undamped double pendulum moving in a plane, which is a chaotic Hamiltonian dynamical system. The double pendulum is a popular example in introductory texts and for physical demonstrations of chaos, where the chaotic dynamics make it difficult for an observer to anticipate when the bottom segment of the pendulum will flip---that is, when it will rotate past the upper segment. Various unpublished notes \cite{heyl_2008, palace, elinson} provide color-coded plots showing how the time until the first flip depends on the initial position of the two segments, and these plots appear to be fractal. There are few results on the double pendulum in the peer-reviewed literature, but Poincar\'e sections, periodic orbits, and Lyapunov exponents have been computed.\cite{stachowiak2006numerical}

The question we ask here is: which stationary initial positions of the pendulum do not lead to flipping of the bottom segment within a chosen time horizon? Relative to past SOS computations of polynomial barrier functions, this more complicated example presents several new challenges. In addition to being chaotic, the system has symmetries and a conserved quantity that may be exploited. Further, the governing ODEs in angular coordinates are not polynomial since they include trigonometric terms, so {special strategies must me employed}. Finally, rather than asking only whether a specified set of initial conditions leads to flipping, we aim to approximate from within the set of all initial positions that do not lead to flipping.

This manuscript proceeds as follows. Section \ref{sec:bfm} presents the barrier function approach and its implementation in the polynomial case as an SOS program. Section \ref{sec:doublependulum} gives the derivation of polynomial ODEs governing the undamped double pendulum. Section \ref{sec:fliptime} defines the flip of the lower segment and the question to be studied, followed in \ref{sec:sosprogram} by the SOS program we solve computationally to construct barrier functions. Results of these computations are reported in section \ref{sec:results}, followed by discussion in section \ref{sec:conclusions}.

\section{\label{sec:formulation}Problem formulation}

\subsection{\label{sec:bfm}Barrier function method}

Consider a well-posed dynamical system
\begin{equation}
    \ddt\mathbf{x}(t) = \mathbf{f}(\mathbf{x}(t),t),
    \label{eq:ode general}
\end{equation}
where $\mathbf f:\mathbb{R}^n\times\R\to\mathbb{R}^n$. Suppose that given functions $g,h:\mathbb{R}^n\to\mathbb{R}$ define the initial set as all points in the phase space $\mathbb R^n$ where $g(\mathbf{x})\leq 0$ and define the avoidance set as all points where $h(\mathbf{x})=0$. (The definition of these sets by a single inequality and equality condition is for clarity of explanation; in general either set can be specified by any number of inequality and equality conditions, and the avoidance set can depend also on $t$.) To show that no trajectories from the initial set enter the avoidance set during the time interval $[0,T]$, it suffices to construct a continuously differentiable barrier function $V:\mathbb R^n\times\mathbb R\to\mathbb R$ such that $V(\mathbf{x},t)$ satisfies the pointwise inequalities

\begin{subequations}
\label{eq:V cond}
\begin{align}
    \tfrac{\partial V}{\partial t}(\x,t) + \f(\x,t)\cdot \nabla_\x V(\x,t) &\geq 0 ~\text{ when }~ t(T-t)\ge0, \label{eq:nonincreasing} \\
    V(\x,0) &> 0 ~\text{ when }~ g(\mathbf{x})\ge0, \label{eq:init general}\\
    V(\x,t) &\leq 0 ~\text{ when }~ h(\mathbf{x}) = 0,~t(T-t)\ge0 \label{eq:avoidanceset}
\end{align}
\end{subequations}

The left-hand expression in \eqref{eq:nonincreasing} can be evaluated for any $V(\x,t)$ without knowing any trajectories $\x(t)$ of the ODE, but by the chain rule it agrees everywhere with $\ddt V(\x(t),t)$, the total time derivative of $V$ along trajectories. Thus, the first constraint means that $V$ is nondecreasing along all trajectories up to time $T$, while \eqref{eq:init general}--\eqref{eq:avoidanceset} mean $V$ is positive on the initial set and nonpositive on the avoidance set. Any $V$ satisfying all three conditions is a barrier function certifying avoidance.

In the case where the ODE right-hand side $\f(\x,t)$ is polynomial in the components of $\x$ and in $t$, and $V(\x,t)$ is sought from a finite-dimensional space of polynomials, the constraints \eqref{eq:V cond} are pointwise inequalities of polynomials on subsets of $\R^n\times\R$. Assume that the initial and avoidance sets are semialgebraic---i.e., they can be specified by polynomial inequalities or equalities; for the constraints in \eqref{eq:V cond} this means that $g(\x)$ and $h(\x)$ are polynomials also. There is a standard way to strengthen the pointwise polynomial inequalities in \eqref{eq:V cond} into tractable SOS conditions. For concreteness consider the constraint \eqref{eq:avoidanceset}. We do not want to simply require $-V(\x,t)$ to be SOS because this would imply $V(\x,t)\le0$ on all of $\R^n\times\R$, which {contradicts \eqref{eq:init general}}. The inequality of \eqref{eq:avoidanceset} only needs to be enforced on the semialgebraic set where $t(T-t)\ge0$ and $h(\x)=0$, and a sufficient condition for this is that
\begin{equation}
    - V -  t(T-t) \sigma(\mathbf{x},t)- h(\mathbf{x})\rho(\mathbf{x},t) ~ \text{ is SOS},
\label{eq:s-proc}
\end{equation}
where $\sigma(\x,t)$ is an SOS polynomial but $\rho(\x,t)$ can be any polynomial. Conditions \eqref{eq:nonincreasing}--\eqref{eq:init general} can be replaced with SOS sufficient conditions in an analogous way, resulting in a problem where $V(\x,t)$ and all ancillary functions like $\sigma(\x,t)$ and $\rho(\x,t)$ are sought from chosen finite-dimensional spaces of polynomials, subject to various SOS constraints in which the tunable coefficients of these polynomials appear linearly. This problem is an instance of an SOS program---a type of convex optimization problem with SOS constraints, where tunable variables appear linearly in the constraints and in the optimization objective (if there is one). Here there is no optimization objective since we only need to verify that there exist polynomials satisfying the constraints. 

The above approach for polynomial $\f(\x,t)$ can be generalized to $\f(\x,t)$ that are rational functions---i.e., ratios of polynomials---with sign-definite denominators. In this case the rational inequality \eqref{eq:nonincreasing} can be multiplied by the denominators to obtain a polynomial inequality, as in section~\ref{sec:sosprogram} below.

\subsection{\label{sec:doublependulum}Rational equations governing the double pendulum}

\begin{figure}[t!]
 \centering
 \includegraphics{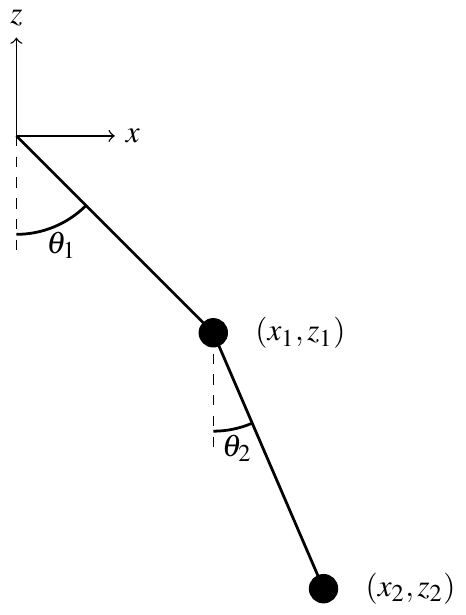}
 \caption{Coordinate system for the double simple pendulum.}
 \label{fig:pendulum}
\end{figure}

The undamped double simple pendulum that we consider consists of two unit masses connected by massless rods of unit length; cf.\ figure~\ref{fig:pendulum}. To formulate the governing ODEs we seek coordinates that lead to a first-order system with a right-hand side that is polynomial, or at least rational, so that we can compute polynomial barrier functions by the SOS approach described above.

We derive our preferred governing equations in several steps, starting with the Cartesian coordinates $(x_i,z_i)$ for the position of each mass. The Lagrangian function is
\begin{align}
\begin{split}
\label{eq:L}
\mathcal{L} = &\frac{1}{2}\left(\dot{x_1}^2+\dot{z_1}^2+\dot{x_2}^2+\dot{z_2}^2\right)-z_1-z_2 \\
              &-\frac{1}{2}\tau_1\left(x_1^2+z_1^2-1\right) \\
              &-\frac{1}{2}\tau_2\left((x_2-x_1)^2+(z_2-z_1)^2-1\right),
\end{split}
\end{align}
where dots denote time derivatives. The variables $\tau_1$ and $\tau_2$ are Langrange multipliers that constrain the length of the rods, but they also are proportional to the tension in the rods. The Euler--Lagrange equations of \eqref{eq:L} give a governing system of ODEs and algebraic constraints,
\begin{align}
\label{eq:odes xz}
\begin{split}
        \ddot{x_1}&=-\tau_1x_1+\tau_2(x_2-x_1),\\
        \ddot{x_2}&=-\tau_2(x_2-x_1),\\
        \ddot{z_1}&=-1-\tau_1z_1+\tau_2(z_2-z_1),\\
        \ddot{z_2}&=-1-\tau_2(z_2-z_1),\\
        1 &= x_1^2+z_1^2, \quad
        1 = (x_2-x_1)^2+(z_2-z_1)^2.
\end{split}
\end{align}
The $\tau_i$ are defined implicitly by these equations through the preservation of the constraints. Although one could introduce additional variables to rewrite this second-order system as a first-order system whose ODEs take the standard form \eqref{eq:ode general} with polynomial $\f(\x)$, we instead change to angular coordinates because this requires fewer variables and simplifies the definition of a pendulum `flip' in section~\ref{sec:fliptime}. Let the offset of each pendulum from vertical be denoted by $\theta_i$ (cf.\ figure~\ref{fig:pendulum}) with angular velocity $\omega_i$, so \eqref{eq:odes xz} becomes
\begin{align}
\begin{split}
\label{eq:trig}
       \dot{\theta_1}&=\omega_1, \qquad \dot{\theta_2}=\omega_2,\\
       \dot{\omega_1}&=\tau_2\sin(\theta_2-\theta_1)-\sin\theta_1,\\
       \dot{\omega_2}&=-\tau_1\sin(\theta_2-\theta_1).
\end{split}
\end{align}
The two constraints for the lengths of the rods are now satisfied automatically, but in reducing the four second-order equations of \eqref{eq:odes xz} to the system above, we additionally find the two constraints
\begin{align}
\begin{split}
\label{eq:trigconst}
        0&=\tau_1-\cos(\theta_2-\theta_1)\tau_2-\omega_1^2-\cos\theta_1,\\
        0&=2\tau_2-\cos(\theta_2-\theta_1) \tau_1-\omega_2^2.
\end{split}
\end{align}
Yet another set of governing equations can be derived by writing the Lagrangian \eqref{eq:L} in terms of angular coordinates, in which case the corresponding Euler--Lagrange equations have more complicated trigonometric expressions than \eqref{eq:trig}, however we choose \eqref{eq:trig} because it can be transformed into a first-order rational system more easily.

To transform \eqref{eq:trig} into a system whose right-hand side is a rational function of the coordinates, we introduce the new variables $c_1=\cos\theta_1$, $s_1=\sin\theta_1$, $c_2=\cos(\theta_2-\theta_1)$, and $s_2=\sin(\theta_2-\theta_1)$. (These definitions of $c_2$ and $s_2$ work better than the more obvious choices $\cos\theta_2$ and $\sin\theta_2$, which would lead to an ODE system of higher polynomial degree.) 
The constraints \eqref{eq:trigconst} give the $\tau_i$ in terms of the $c_i$ and $s_i$ as
\begin{align}
\label{eq:tensions}
\tau_1&=\frac{2\omega_1^2+c_2\omega_2^2+2c_1}{1+s_2^2}, &
\tau_2&=\frac{\omega_2^2+c_2\omega_1^2+c_1c_2}{1+s_2^2}.
\end{align}
Finally, we arrive at a first-order rational system {that includes the evolution equations for $\omega_i$ as well as the evolution equations found by differentiating $c_i$ and $s_i$, which are needed to close the ODE system:}
\begin{equation}
\begin{aligned}
\dot{\omega_1}&=\frac{s_2\omega_2^2+s_2c_2\omega_1^2+c_1c_2s_2 - s_1-s_1s_2^2}{1+s_2^2},\\
\dot{\omega_2}&=-\frac{s_2\omega_2^2+s_2c_2\omega_1^2+c_1c_2s_2}{1+s_2^2},\\
\dot{c_1}&=-\omega_1s_1, &
\dot{s_1}&=\omega_1c_1,\\
\dot{c_2}&=-(\omega_2-\omega_1)s_2, &
\dot{s_2}&=(\omega_2-\omega_1)c_2,
\end{aligned}
\label{eq:ourvariables}
\end{equation}
where all solutions of interest satisfy the two constraints
\begin{equation}
\label{eq:ourconstraints}
\begin{aligned}
    h_i(c_i,s_i):=c_i^2+s_i^2-1 =0,\quad i=1,2,
\end{aligned}
\end{equation}
which fixes the correct magnitudes for the $c_i$ and $s_i$. Defining sine and cosine variables to replace the trigonometric terms in~\eqref{eq:trig} with polynomial terms in \eqref{eq:ourvariables} is a common strategy for applying SOS methods to trigonometric systems.\cite{Papachristodoulou2005b} It works well here, so we do not pursue alternatives such as including trigonometric functions in the basis vector used in translating the problem into a semidefinite program. Another possibility would be to eliminate the sine and cosine variables in favor of a cotagent, which gives a system of equations with smaller dimension but higher polynomial degree.

The total energy $E$ of the system is a conserved quantity that we make use of below. In the variables of \eqref{eq:ourvariables} and with a constant chosen so that the minimum energy is zero,
\begin{equation}
\label{eq:energy}
E = 3-2c_1-c_2c_1+s_2s_1+\omega_1^2+\tfrac{1}{2}\omega_2^2+c_2\omega_1\omega_2.
\end{equation}

\subsection{\label{sec:fliptime}Definition of a flip}

We define a `flip' of the double pendulum as occurring when the two rods cross each other ($\theta_2-\theta_1 = \pm \pi$). Other ways to define a flip include when either of the rods becomes vertical \cite{heyl_2008, elinson} or when the outer rod becomes vertical\cite{palace}. Any of these definitions serves our current purpose, which is to have a property of a chaotic system that is easy to characterize as a subset of phase space, but our chosen definition is especially simple to state in the variables of \eqref{eq:ourvariables} as $c_2=-1$.

\begin{figure}[t]
    \centering
    \includegraphics[width=\columnwidth]{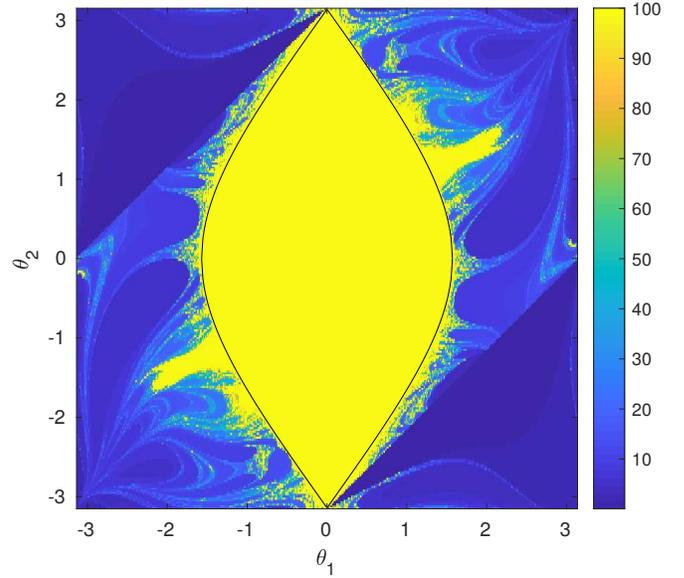}
    \caption{Time until first flip (truncated at $t=100$) for all stationary initial conditions, as computed by numerical integration from a finite-resolution grid of the \mbox{$\theta_1$--$\theta_2$} plane. Inside the almond-shaped curve in the middle, flipping is energetically impossible.}
    \label{fig:flipfractal}
\end{figure}

Figure \ref{fig:flipfractal} shows the time-to-flip for initial conditions that are stationary---i.e., $\omega_1,\omega_2=0$---with different combinations of the initial $\theta_1$ and $\theta_2$. To make this image, we covered the \mbox{$\theta_1$--$\theta_2$} plane with a finite-resolution grid and numerically integrating \eqref{eq:trig}--\eqref{eq:trigconst} starting from each grid point until either a flip occurred or $t=100$ was reached. (We used MATLAB's \texttt{ode45} function with an absolute tolerance of $10^{-10}$ and a relative tolerance of $10^{-7}$.) Our ability to so characterize time-to-flip using numerical integration relies on the initial set of interest being low-dimensional since the cost of covering it with a grid grows exponentially with its dimension. This is why we limit our example to stationary initial conditions, in which case the initial set is two-dimensional. The computational construction of barrier functions reported below is no harder for full-dimensional initial sets---indeed this is a main reason our SOS approach can be useful---but it would be much harder to verify and visualize the same results using numerical integration, which we want to do here to study the effectiveness of the SOS approach.

The rotational symmetry of figure~\ref{fig:flipfractal} reflects the fact that the dynamics and the definition of a flip both are invariant under negation of all four variables in \eqref{eq:trig}. The corresponding symmetry in the enlarged system \eqref{eq:ourvariables} negates the $\omega_i$ and $s_i$ but not the $c_i$. The time-to-flip is zero on the diagonal lines where $\theta_2-\theta_1 = \pm \pi$ and is infinite inside the almond-shaped curve that corresponds to the minimum-energy flipped position $(\theta_1,\theta_2)=(0,\pm\pi)$. The level set curves are simple shapes for short time-to-flip values, but as time increases they become complicated and appear to approach fractal sets.

\subsection{\label{sec:sosprogram}Sum-of-squares program}

A barrier function $V(\x,t)$ for the double pendulum in the variables of \eqref{eq:trig} must satisfy the three left-hand inequalities in \eqref{eq:V cond} with $\x=(\omega_1,\omega_2,c_1,c_2,s_1,s_2)$ being the state vector of the rational ODE system \eqref{eq:ourvariables} and $\f(\x)$ being its right-hand vector. However, the conditions defining the sets where these pointwise inequalities must hold are more complicated than those on the right-hand side of \eqref{eq:V cond}. The $V$ inequalities need to hold only on trajectories that satisfy the two $h_i=0$ constraints \eqref{eq:ourconstraints} and whose energy \eqref{eq:energy} is less than some upper bound $E_u$ on the energies of the initial set we consider. We define
\begin{align}
h_3&:=c_2+1,&
g_1&:= t(T-t), &
g_2&:= E_u-E,
\end{align}
so that $h_3=0$ corresponds to an exactly flipped pendulum---i.e., our avoidance set---while $g_1\ge0$ and $g_2\ge0$ on the time interval $[0,T]$ and energy interval $(-\infty,E_u]$, respectively. In the next subsection we introduce one more polynomial, $g_3(c_1,c_2,s_1,s_2)$, such that each initial set of interest is defined by $g_3\ge0$ along with $\omega_1,\omega_2=0$. To show that no trajectories from an initial set enter the avoidance set up to time $T$, it suffices to find a $V$ satisfying
\begin{subequations}
\label{eq:V cond 2}
\begin{align}
    \tfrac{\partial V}{\partial t} + \f\cdot \nabla_\x V&\geq 0 ~\text{ when }~ h_1,h_2=0, ~g_1,g_2\ge0, \label{eq:V cond 2 first} \\
    V|_{t=0} &\ge \epsilon ~\text{ when }~ \omega_1,\omega_2=0,~g_3\ge0,\\
    V &\leq 0 ~\text{ when }~ h_1,h_2,h_3 =0,~g_1,g_2\ge0,
\end{align}
\end{subequations}
where $\epsilon$ is {a positive constant that ensures} strict positivity of $V(\x,0)$ on the initial set.
{In theory the value of $\epsilon$ makes no difference since a $V$ that satisfies \eqref{eq:V cond 2} for one positive $\epsilon$ can be rescaled to satisfy \eqref{eq:V cond 2} for any other value of $\epsilon$. In practice, however, we find better numerical performance with small values of $\epsilon$, so we fix $\epsilon=10^{-3}$ here.}

The rational terms on the right-hand side $\f(\x)$ of the governing system \eqref{eq:ourvariables} have positive denominator $D:=1+s_2^2$, so multiplying the rational inequality in \eqref{eq:V cond 2 first} by $D$ gives an equivalent polynomial inequality. Then all three inequalities in \eqref{eq:V cond 2} are polynomial and can be strengthened into SOS conditions using the approach illustrated by \eqref{eq:s-proc}. In particular, it suffices for the three expressions
\begin{subequations}
\label{eq:fullproblem}
\begin{align}
&D\tfrac{\partial V}{\partial t}+D\mathbf{f}\cdot\nabla_\x V- D h_1 \rho_1 - D h_2 \rho_2 -D g_1\sigma_1 - D g_2 \sigma_2,\label{eq:fullproblemfirst}\\
    &V|_{t=0}-\epsilon-\omega_1^2 \rho_3-\omega_2^2\rho_4-g_3\sigma_3,\label{eq:initial}\\
\begin{split}
    &-V - h_1\rho_5 - h_2 \rho_6- h_3\rho_7-g_1\sigma_4 - g_2\sigma_5,
\end{split}\label{eq:target}
\end{align}
\end{subequations}
to be SOS, where all five $\sigma_i$ are SOS also but the $\rho_i$ polynomials need not be. The tunable polynomials $\rho_i$ and $\sigma_i$ depend on $\x$ in \eqref{eq:initial} and on $(\x,t)$ in (\ref{eq:fullproblem}a,c). In \eqref{eq:initial} the $\omega_i$ are squared so that the full expression shares the symmetry $(\omega_i,s_i)\mapsto-(\omega_i,s_i)$ of the ODE system \eqref{eq:ourvariables}, provided symmetries are also imposed on the $\rho_i$ and $\sigma_i$ as described in the next subsection. {Note that because the initial set is simple to characterize by the condition $\omega_i=0$, one could avoid the need for $\rho_3$ and $\rho_4$ by substituting these $\omega_i$ values into the other terms in \eqref{eq:initial} and enforcing nonnegativity without the $\omega_i$ variables; we use the slightly more expensive condition \eqref{eq:fullproblem} that fits within the general framework of \eqref{sec:bfm}.} In total there are eight SOS constraints, and in these the coefficients of the tunable polynomials ($V$, $\rho_i$, $\sigma_i$) appear linearly. Once we specify finite-dimensional spaces from which to seek each tunable polynomial, this constitutes an SOS program.

\subsection{\label{sec:initialsets}Initial sets}

It remains to choose the polynomials $g_3(c_1,c_2,s_1,s_2)$ that, along with the stationarity conditions $\omega_1,\omega_2=0$, will define the initial sets of interest. Our aim is to fix a final time $T$ and approximate (from within) the complicated set of stationary positions $(\theta_1,\theta_2)$ that do not lead to a flip during the time interval $[0,T]$. It is necessary to fix $g_3$ before solving the SOS program, so the initial set must be fixed rather than optimized in some way. The reason that $g_3$ cannot be tunable is that it multiplies the tunable polynomial $\sigma_3$ in \eqref{eq:initial}, and tunable coefficients can appear only linearly in SOS constraints in order for the problem to remain convex. Thus, we consider various $g_3$ that define many different initial sets, and together these initial sets cover the \mbox{$\theta_1$--$\theta_2$} plane in oval shapes. For each initial set, we solve an SOS program to seek a barrier function $V$. If successful, then no trajectories from this initial set flip within time $[0,T]$. If unsuccessful, such a flip may or may not occur for at least one trajectory in the initial set, and the attempt to find a barrier function can be repeated for a smaller initial set and/or with the tunable polynomials ($V$, $\rho_i$, $\sigma_i$) sought from larger spaces.

The initial sets that we choose are balls in the $(c_1,c_2,s_1,s_2)$ variables, combined with their counterparts under the system's symmetry. These sets correspond to nonnegativity of the quartic polynomial
\begin{equation}
\begin{aligned}
g_3 :=& \left[(c_1-c_1^0)^2 + (c_2-c_2^0)^2 + (s_1-s_1^0)^2 + (s_2-s_2^0)^2-R^2\right] \\
&\times
\left[(c_1-c_1^0)^2 + (c_2-c_2^0)^2 + (s_1+s_1^0)^2 + (s_2+s_2^0)^2-R^2\right],
\end{aligned}
\end{equation}
where numerical values for the superscripted quantities must be fixed to define the ball's center, and $R$ defines the radius. In terms of the angular position variables $(\theta_1,\theta_2)$, the condition $g_3\ge0$ corresponds to an oval shape oriented diagonally in the \mbox{$\theta_1$--$\theta_2$} plane, along with its symmetric counterpart on which both $\theta_i$ are negated. The reason that $g_3$ has been chosen to specify balls in symmetry-related pairs is so that the barrier function constraints \eqref{eq:V cond 2} all remain invariant under the system's symmetry, which in terms of the variables of \eqref{eq:ourvariables} is $(\omega_i,s_i)\mapsto-(\omega_i,s_i)$. This lets us save significant computational cost by imposing the same invariance on the tunable polynomials $(V,\rho_i,\sigma_i)$. This restriction has no effect on the existence of a barrier function since existence of non-symmetric tunable polynomials implies existence of symmetrized versions, as can be shown by an argument analogous to theorem 2 in \citet{lakshmi2020finding}.

To produce a collection of balls in the $(c_1,s_1,c_2,s_2)$ variables that correspond to a covering of the $\theta_1-\theta_2$ plane by ovals, we first divide the region where $(\theta_1,\theta_2)\in[0,\pi]\times[-\pi,\pi]$ into a grid of $7\times 15$ rectangles. For each rectangle, we find the smallest ball in $(c_1,s_1,c_2,s_2)$ whose corresponding oval in $(\theta_1,\theta_2)$ covers the rectangle. (This is done numerically by solving a simple SOS program to find the minimum such $R$ for a ball centered on the same point as the rectangle.) Having chosen the initial set, we then find an upper bound $E_u$ on energies in this set by solving another simple SOS program. To verify that no trajectories from the initial set flip during $[0,T]$, we solve the SOS program described in section~\ref{sec:sosprogram}, where the eight expressions required to be SOS are \eqref{eq:fullproblemfirst}--\eqref{eq:target} and $\sigma_1,\ldots,\sigma_5$. If this verification fails, we subdivide the rectangle into four and repeat the process, attempting to verify that no flip occurs for smaller initial sets covering each smaller rectangle. This subdivision can be repeated to whatever resolution is desired. In our computations we carry out two levels of subdivision over $[0,\pi]\times[-\pi,\pi]$. We also study the smaller region $[2\pi/3,\pi]\times[2\pi/3,4\pi/3]$ with higher resolution, dividing it first into $7\times 15$ rectangles and attempting up to three levels of subdivision.

{This iterative process allows us to gradually map out much more complicated shapes than would be possible with SOS methods that do not decompose phase space. The method of Ref.\ \citenum{Henrion2014}, for instance, would require extremely high polynomial degrees to capture the fractal-like patterns observed in the double pendulum system. The adaptation of the latter method by Ref.\ \citenum{Cibulka2022} would improve performance but would likely still require polynomial degrees for which computations are either intractable or extremely expensive. On the other hand, the methods of Refs.\ \citenum{Henrion2014} and \citenum{Cibulka2022} do have theoretical convergence guarantees that are not yet proved for our approach.}

\section{\label{sec:results}Computational results}

To test our computational approach, we use the procedure described at the end of the preceding section to find oval sets of initial conditions in the \mbox{$\theta_1$--$\theta_2$} plane that do not lead to flipping prior to time $T=6$. For each initial set we solve the SOS program described in section~\ref{sec:sosprogram} with tunable polynomials ($V$, $\rho_i$, $\sigma_i$) that include all monomials invariant under $(\omega_1,\omega_2,s_1,s_2)\mapsto-(\omega_1,\omega_2,s_1,s_2)$ whose degree in $t$ and total degree in the components of $\x$ are at most 6. In some cases a barrier function cannot be found using these polynomials but can be found if the polynomial degrees are raised or if the radius of the initial set is made smaller. There is a tradeoff in computation between using higher-degree polynomials, which raises the cost of each SOS program, and covering the \mbox{$\theta_1$--$\theta_2$} plane using smaller initial sets, which requires solving more SOS programs. The spaces from which we seek each tunable polynomial represent a compromise in this tradeoff; omitting degree-6 terms makes it much harder to find barrier functions, and adding higher-degree terms raises the computational cost quickly.

In order to improve numerical conditioning in the solution of the SOS programs, we found it necessary to follow a heuristic of previous authors \cite{Henrion2014, Goluskin2018} in which variables are rescaled such that all $(\x,t)$ satisfying the right-hand conditions in \eqref{eq:V cond 2} lie within the hypercube $[-1,1]^7$. Already $|c_i|,|s_i|\le1$ due to the $h_1,h_2=0$ constraints, which reflect that fact that these variables are sines and cosines. Initial conditions of interest are stationary and so have energy $E\le6$, which implies $|\omega_1|\leq2\sqrt{3}$ and $|\omega_2|\leq2\sqrt{6}$.
Therefore, when we formulate SOS programs for numerical solution, we do so in terms of the scaled variables $\widetilde{\omega}_1=\omega_1/2\sqrt{3}$, $\widetilde{\omega}_2=\omega_2/2\sqrt{6}$, and $\widetilde{t}=t/T$. 

Numerical solution of each SOS program was implemented in MATLAB using the parser YALMIP\cite{lofberg2004yalmip} (version R20200116) to define the SOS program and translate it into an equivalent semidefinite program. This semidefinite program was solved using Mosek \cite{Andersen2000} (version 9.1)
in parallel on 18 Intel Skylake cores with shared memory. Approximately 26GB of shared memory and 1.5 hours of wall time were required by Mosek in each case. Initial sets closer to the boundary of the non-flipping region typically required more iterations of Mosek's interior-point algorithm to reach a given convergence tolerance. 

\begin{figure}[t]
\centering
\begin{overpic}[width=\columnwidth]{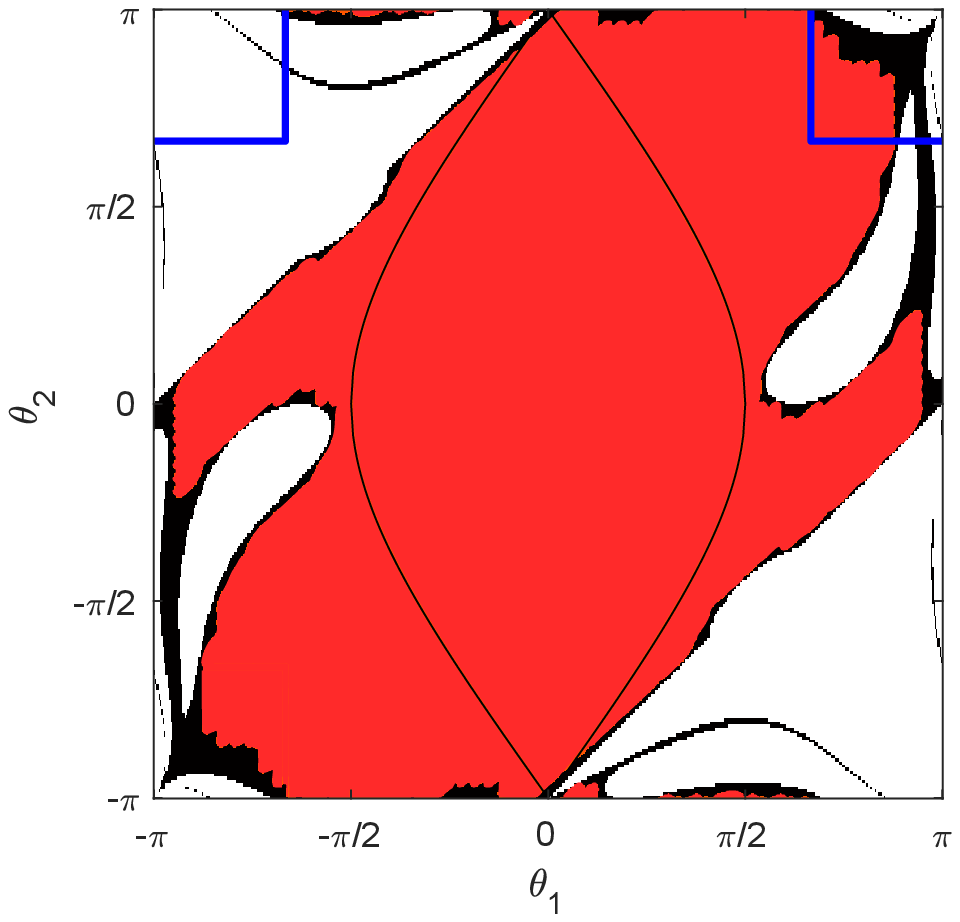}
    \put (0,90) {(a)}
\end{overpic}
\begin{overpic}[width=\columnwidth]{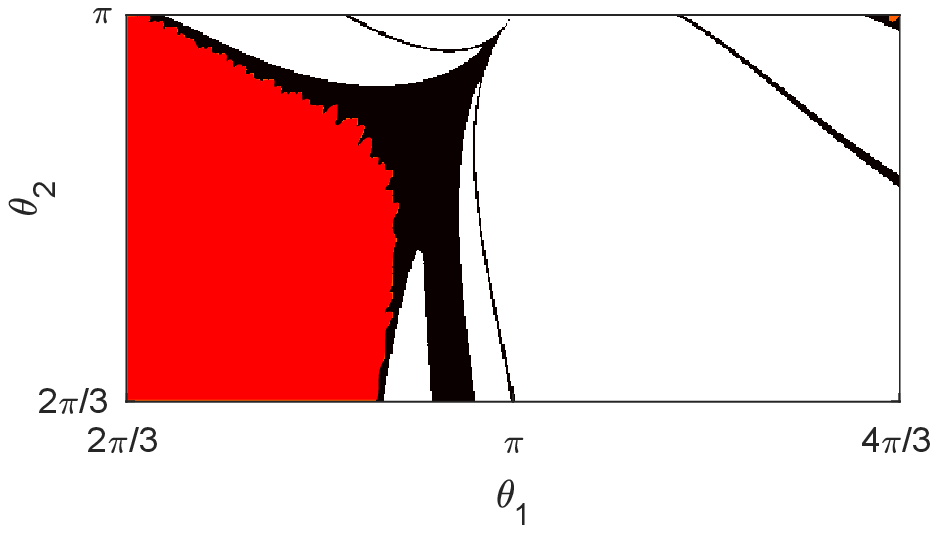}
    \put (0,50) {(b)}
\end{overpic}
\caption{Initial positions $(\theta_1,\theta_2)$ for which a stationary double pendulum will not flip before time $T=6$, as approximated using numerical integration (black) and as verified on a subset of this region using SOS computations (red). Initial positions in the white regions lead to flips at $t\le6$. Panel (a) shows the full periodic space of initial positions, while panel (b) shows the region spanning the indicated corners of panel (a).}
\label{fig:T6}
\end{figure}

Figure~\ref{fig:T6} shows the set of initial positions $(\theta_1,\theta_2)$ for which the pendulum does not flip before $T=6$, as approximated by numerically integration from a fine grid of initial conditions (larger black region) and as verified by constructing barrier functions using SOS computations (smaller red region). Although the region found by numerical integration is larger, its resolution is finite. In black regions where flipping has not been ruled out by barrier functions, increasing the resolution might reveal white regions with finer structure---possibly measure-zero structure that would not be found at any finite resolution. In the red region, on the other hand, flipping of the pendulum is ruled out on the entirety of each covering oval. In this sense the barrier function approach provides a more certain result, at the expense of giving a somewhat conservative inner approximation of the non-flipping set. The red region in figure~\ref{fig:T6} could be enlarged by seeking smaller ovals or raising the degrees of tunable polynomials in each SOS program, but it will always be smaller than the black region. Figure~\ref{fig:T6}b shows the results of the method being performed on the smaller region spanning the corners of figure~\ref{fig:T6}a, with a third subdivision step. A clear boundary is visible beyond which barrier functions have not ruled out flipping, despite numerical integration suggesting its impossibility.

\section{\label{sec:conclusions}Conclusions}

The construction of polynomial barrier functions by sum-of-squares computations presented in this work is, as far as we know, the first such application to a chaotic system. In the course of applying these polynomial optimization methods to the double pendulum, we have illustrated how to transform a trigonometric ODE system into a larger rational systems with constraints, as well as how to take advantage of a conserved quantity and a discrete symmetry. By computing barrier functions we found many sets of initial positions of a stationary double pendulum that do not lead to flipping within a specified time. The union of these sets is an inner approximation to the complicated set of all non-flipping initial conditions. By constructing barrier functions of fairly low polynomial degree, we were able to verify most of the non-flipping set but not regions in its fine filaments or very near its boundary. 

The total computational cost of solving our SOS programs was much higher than the cost of the brute force approach in which numerical integration was carried out for a fine grid of initial conditions over the set of interest. However, this is partly because we designed our example to make numerical integration easier: restricting to stationary initial conditions reduces the possible initial states from a four-dimensional set to a two-dimensional one, and far fewer points are needed to cover the latter in a fine grid. In applications where the dimension of the initial set is higher, the computational cost of numerical integration over a fine grid may be prohibitive while the barrier function approach remains tractable, {provided that the sets used to cover the initial set are not too small.} Additionally, the cost of computing barrier functions may be warranted when the fact that trajectories from an initial set do not enter an avoidance set must be verified with the rigor of a computer-assisted proof; after an SOS program is reformulated as a semidefinite program, the numerical solution of the latter can be made rigorous using interval arithmetic \cite{Jansson2006, Goluskin2018}.
The ability of SOS computations to produce barrier functions in examples with higher-dimensional initial sets, or when augmented with interval arithmetic, remains to be explored in future work.

\begin{acknowledgments}
This work is dedicated to the memory of Charlie Doering. This paper began as a project in the Geophysical Fluid Dynamics summer program at the Woods Hole Oceanographic Institution. Charlie played an important part in the early stages of this work and provided much encouragement and insight throughout. He will be sorely missed. We also thank P.\ J.\ Morrison and G.\ Fantuzzi for helpful discussions{, and an anonymous reviewer for useful comments.} DG was supported by the Canadian NSERC Discovery Grants Program awards RGPIN-2018-04263, RGPAS-2018-522657 \& DGECR-2018-0037.
\end{acknowledgments}

\section*{Data Availability Statement}
The data that support the findings of this study are available from the corresponding author upon reasonable request.

\appendix

\bibliography{pgv}

\end{document}